\documentstyle[12pt]{article}

\def\Lms{\Lambda_{\overline{MS}}}
\def\Llo{\Lambda_{LO}}

\begin{document}
\thispagestyle{empty}
\begin{flushright}
 MPI-PhT/94-60 \\
 TUM-T31-75/94 \\
 September 1994
\end{flushright}
\vskip1truecm
\centerline{\Large\bf  QCD Factors $a_1$ and $a_2$ Beyond Leading}
\centerline{\Large\bf Logarithms versus Factorization in}
\centerline{\Large\bf Non-leptonic Heavy Meson Decays
   \footnote[1]{\noindent
   Supported by the German
   Bundesministerium f\"ur Forschung und Technologie under contract
   06 TM 732 and by the CEC science project SC1--CT91--0729.}}
\vskip1truecm
\centerline{\sc Andrzej J. Buras}
\bigskip
\centerline{\sl Technische Universit\"at M\"unchen, Physik Department}
\centerline{\sl D-85748 Garching, Germany}
\vskip0.6truecm
\centerline{\sl Max-Planck-Institut f\"ur Physik}
\centerline{\sl  -- Werner-Heisenberg-Institut --}
\centerline{\sl F\"ohringer Ring 6, D-80805 M\"unchen, Germany}

\vskip1truecm
\centerline{\bf Abstract}
We calculate the QCD factors $a_1$ and $a_2$, entering the tests of
factorization in non-leptonic heavy meson decays, beyond the leading
logarithmic approximation in three renormalization schemes
(NDR, HV, DRED). We
investigate their $\mu$--dependence and the renormalization
scheme dependence. We point out that $a_1$ for B-decays depends
very weakly on
$\Lms$, $\mu$ and the choice of the renormalization scheme. For
$\Lms^{(5)}=225\pm85~MeV$ and $4~GeV\leq\mu\leq 8~GeV$ we find
$a_1=1.01\pm0.02$ in accordance
with phenomenology. The $\Lms$, $\mu$ and scheme
dependences of $a_2$ are on the other hand sizable. Interestingly,
for the NDR scheme we find $a_2^{NDR}=0.20\pm0.05$,
in the ball park of recent phenomenological results and
substantially larger than leading order estimates.
However $a_2^{HV}=0.16\pm0.05$ and $a_2^{DRED}=0.15\pm0.05$.
Implications of these
findings for the tests of factorization in B-decays
and D-decays are critically discussed.

\vfill
\newpage

\pagenumbering{arabic}
\section{Introduction}
In the factorization approach to non-leptonic meson decays
\cite{FEYNMAN,STECHF} one can
distinguish three classes of decays for which the amplitudes have the
following general structure \cite{BAUER,NEUBERT}:
\begin{equation}\label{1}
A_{\rm I}=\frac{G_F}{\sqrt{2}} V_{CKM}a_1(\mu)\langle O_1\rangle_F
\qquad {\rm (Class~I)}
\end{equation}
\begin{equation}\label{2}
A_{\rm II}=\frac{G_F}{\sqrt{2}} V_{CKM}a_2(\mu)\langle O_2\rangle_F
\qquad {\rm (Class~II)}
\end{equation}
\begin{equation}\label{3}
A_{\rm III}=
\frac{G_F}{\sqrt{2}} V_{CKM}[a_1(\mu)+x a_2(\mu)]\langle O_1\rangle_F
 \qquad {\rm (Class~III)}
\end{equation}
Here $V_{CKM}$ denotes symbolically the CKM factor characteristic for a
given decay. $O_1$ and $O_2$ are local four quark operators present in
the relevant effective hamiltonian, $\langle O_i\rangle_F$ are
the hadronic matrix
elements of these operators given as products of matrix elements of
quark currents and $x$ is a non-perturbative factor equal to unity in
the flavour symmetry limit. Finally $a_i(\mu)$ are $QCD$ factors which
are the main subject of this paper.

As an example consider the decay $\bar B^0\to D^+\pi^-$. Then the
relevant effective hamiltonian is given by
\begin{equation}\label{4}
H_{eff}=\frac{G_F}{\sqrt{2}}V_{cb}V_{ud}^{*}
\lbrack C_1(\mu) O_1+C_2(\mu)O_2 \rbrack
\end{equation}
where
\begin{equation}\label{5}
O_1=(\bar d_i u_i)_{V-A} (\bar c_j b_j)_{V-A}
\qquad
O_2=(\bar d_i u_j)_{V-A} (\bar c_j b_i)_{V-A}
\end{equation}
with $(i,j=1,2,3)$ denoting colour indices and $V-A$ referring to
$\gamma_\mu (1-\gamma_5)$. $C_1(\mu)$ and $C_2(\mu)$ are short distance
Wilson coefficients computed at the renormalization scale $\mu=O(m_b)$.
We will neglect the contributions of penguin operators since their
Wilson coefficients are numerically very small as compared to $C_{1,2}$
\cite{BJL,ROME}. Exceptions are CP-violating decays and rare decays which
are beyond the scope of this paper.
Note that we use here the labeling of the operators as given in
\cite{BAUER,NEUBERT} which differs from \cite{BJL,ROME} by the interchange
$1\leftrightarrow 2$.
$C_i$ and $a_i$ are related as follows:
\begin{equation}\label{6}
a_1(\mu)=C_1(\mu)+\frac{1}{N} C_2(\mu) \qquad
a_2(\mu)=C_2(\mu)+\frac{1}{N} C_1(\mu)
\end{equation}
where $N$ is the number of colours. We will set $N=3$ in what follows.

Application of the factorization method gives
\begin{equation}\label{7}
A(\bar B^0\to D^+\pi^-)=\frac{G_F}{\sqrt{2}}V_{cb}V_{ud}^{*}
a_1(\mu)\langle\pi^-\mid(\bar d_i u_i)_{V-A}\mid 0\rangle
\langle D^+\mid (\bar c_j b_j)_{V-A}\mid \bar B^0\rangle
\end{equation}
where $\langle D^+\pi^-\mid O_1 \mid \bar B^0\rangle$ has been factored
 into two
quark current matrix elements and the second term in $a_1(\mu)$ represents
the contribution of the operator $O_2$ in the factorization approach.

Other decays can be handled in a similar manner \cite{NEUBERT}.
 Although the flavour
structure of the corresponding local operators changes,
the colour structure
and the coefficients $C_i(\mu)$ remain unchanged. For instance
$\bar B^0\to \bar K^0\psi$ and $B^-\to D^0 K^-$ belong to class II and
III respectively. Finally a similar procedure  can be applied to
D-decays with the coefficients $C_i$ evaluated at $\mu=O(m_c)$.
Once the matrix elements have been expressed in terms of various meson
decay constants and generally model dependent formfactors, predictions
for non-leptonic heavy meson decays can be made.
Moreover relations between non-leptonic and semi-leptonic decays can
be found which allow to test factorization in a model independent
manner.

Although the simplicity of this framework is rather appealing,
it is well known that
non-factorizable contributions must be present in the hadronic matrix
elements of the current--current operators $O_1$ and $O_2$ in order
to cancel the $\mu$ dependence of $C_i(\mu)$ or $a_i(\mu)$ so that
the physical amplitudes do not depend on the arbitrary renormalization
scale $\mu$.
$\langle O_i\rangle_F$ being products of matrix elements of
conserved currents
are $\mu$ independent and the cancellation of the $\mu$ dependence
in (\ref{1})-(\ref{3}) does not take place.
Consequently from the point of view of QCD
the factorization approach can be at best correct at a single value
of $\mu$, the  so-called factorization scale $\mu_F$. Although the
approach itself does not provide the value of $\mu_F$, the proponents
of factorization expect $\mu_F=O(m_b)$ and $\mu_F=O(m_c)$ for
B-decays and D-decays respectively.

Here we would like to point out that beyond the leading logarithmic
approximation for $C_i(\mu)$ a new complication arises. As stressed
in \cite{BJLW}, at next to leading level in the renormalization
group improved perturbation theory the coefficients $C_i(\mu)$
depend on the renormalization scheme for operators. Again only
the presence of non-factorizable contributions
 in $\langle O_i\rangle$ can
remove this scheme dependence in the physical amplitudes.
However $\langle O_i\rangle_F$ are renormalization scheme
independent and the factorization approach is of course unable
to tell us whether it works better with an anti-commuting $\gamma_5$
in $D\not=4$ dimensions ( NDR scheme) or with another definition
of $\gamma_5$ such as used in HV (non-anticommuting $\gamma_5$ in
$D\not=4$) or DRED ($\gamma_5$ in $D=4$) schemes.
The renormalization scheme dependence of $a_i$ emphasized here
is rather
annoying from the factorization point of view as it precludes
a unique phenomenological determination of $\mu_F$ as we will
show explicitly below.

On the other hand, arguments have been given
\cite{BJORKEN,DUGAN,NEUBERT} that once $H_{eff}$ in (\ref{4})
has been constructed, factorization could be
approximately true in the case of two-body decays with high
energy release \cite{BJORKEN}, or in certain kinematic regions
\cite{DUGAN}. We will not repeat here these arguments, which
can be found in the original papers as well as in a critical
analysis of various aspects of factorization presented in
\cite{ISGUR}.
Needless to say the issue of factorization does not only
involve the short distance gluon corrections discussed here
but also final state interactions which are discussed in these
papers.

It is difficult to imagine that factorization
can hold even approximately in all circumstances.
In spite of this, it
became fashionable these days to
test this idea,
to some extent, by using a certain set of formfactors to calculate
$ \langle O_i\rangle_F  $ and by making global fits of
the formulae (\ref{1})-(\ref{3})
to the data treating
$ a_1 $ and $ a_2 $ as free independent parameters. The most recent
analyses of this type give for non-leptonic two-body B-decays
\cite{DEANDREA}-\cite{KAMAL}
\begin{equation}\label{8}
a_1\approx 1.05\pm0.10
\qquad
a_2\approx 0.25\pm0.05
\end{equation}
which is compatible with earlier analyses \cite{NEUBERT,STONE90}.
The new CLEO II data \cite{CLEO} favour
a  {\it positive} value of $a_2$ in contrast to earlier expectations
\cite{BAUER,RUCKL} based on extrapolation from charm decays.
At the level of accuracy of the existing  experimental data and
because of strong model
dependence in the relevant formfactors it is not yet possible
to conclude on the basis of these analyses whether the
factorization approach is a useful approximation in general or not.
It is
certainly  conceivable that factorization may apply better to some
non-leptonic decays than to others
\cite{NEUBERT,BJORKEN,DUGAN,ISGUR,BIGI,RUCKL94}
and using all decays in a global fit may misrepresent the true
situation.

Irrespective of all these reservations let us ask
whether the numerical values in (\ref{8}) agree with
the QCD expectations for
$\mu=0(m_b)?$

A straightforward calculation of $a_i(\mu)$ with $C_i(\mu)$ in the
leading logarithmic approximation  \cite{LEE} gives for $ \mu= 5.0~GeV $
and the QCD scale $\Lambda_{LO} = 225\pm85~MeV$
\begin{equation}\label{9}
a_1^{LO}=1.03\pm0.01\quad \qquad a_2^{LO}=0.10\pm0.02
\end{equation}
Whereas the result for $ a_1 $ is compatible with the experimental findings,
the theoretical value for $ a_2 $ disagrees roughly by a factor of two.
The solution to this problem by dropping the $1/N$ terms in (\ref{6})
suggested in \cite{BAUER} and argued for in \cite{RUCKL,BSQCD,BLOK} gives
$a_1^{LO}=1.12\pm 0.02$ and $a_2^{LO}=-0.27\pm0.03$. Whereas the absolute
magnitudes for $a_i$ are consistent with (\ref{8}), the sign of $a_2$ is
wrong. It
has been remarked in \cite{NEUBERT} that the value of $a_2$ could be
increased by
using (\ref{6}) with $ \mu >> m_b $ .
Indeed as shown in table 1 for $\mu=15-20~GeV$
the calculated values for $a_1$ and $a_2$ are compatible with (\ref{8}).
The large value of $\mu=(3-4)~m_b$ is, however, not really what
the proponents of factorization would expect.

\begin{table}[thb]
\caption{Leading order coefficients
$a_1^{LO}$ and $a_2^{LO}$ for B-decays.}
\begin{center}
\begin{tabular}{|c|c|c||c|c||c|c|}
\hline
& \multicolumn{2}{c||}{$\Llo^{(5)}=140~MeV$} &
  \multicolumn{2}{c||}{$\Llo^{(5)}=225~MeV$} &
  \multicolumn{2}{c| }{$\Llo^{(5)}=310~MeV$} \\
\hline
$\mu [GeV]$ & $a_1$ & $a_2$ & $a_1$ &
$a_2$ & $a_1$ & $a_2$  \\
\hline
\hline
5.0 & 1.024 & 0.124 & 1.030 & 0.099 & 1.035 & 0.078
\\
\hline
10.0 & 1.011 & 0.191 & 1.014 & 0.176 & 1.016 & 0.164
\\
\hline
15.0 & 1.007 & 0.224 & 1.008 & 0.214 & 1.009 & 0.205
\\
\hline
20.0 & 1.004 & 0.246 & 1.005 & 0.238 & 1.006 & 0.231
\\
\hline
\end{tabular}
\end{center}
\end{table}

Yet it should be recalled that in order to address the issue of the value
 of $ \mu $
corresponding to the findings in (\ref{8}) it is mandatory to go beyond the
leading logarithmitic approximation and to include at least
the next-to-leading (NLO)
terms. In particular only then one is able to use meaningfully the value
for $\Lms$ extracted from high energy processes. As an illustration we
have used in (\ref{9}) $\Lambda_{LO}=\Lms$ which is of course rather
 arbitrary.
To our surprise no NLO analysis of $ a_1 $ and $ a_2 $ has been
presented in the literature in spite of the fact that the NLO
corrections to $ C_1 $ and $ C_2 $ have been known for
many years \cite{ALT,WEISZ}.

At this point an important warning should be made. The coefficients
$ C_1 $ and $C_2 $ as given in \cite{ALT,WEISZ} and also in \cite{BJLW}
cannot simply be inserted
into (\ref{6}) as done often in the literature.
As stressed in \cite{BJL} the coefficients
 given in \cite{ALT,WEISZ,BJLW}
 differ from the true coefficients of the operators $O_i$
 by $ O(\alpha_s) $ corrections which
have been included in these papers in order to remove the renormalization
scheme dependence. The only paper which gives the true $ C_1 $ and
$ C_2 $ for B-decays is  ref. \cite{BJL},
where these coefficients have been
given for
the NDR and HV renormalization
schemes.

Now the main topic of ref. \cite{BJL} was the ratio
$ \varepsilon'/\varepsilon $. Consequently
the full set of ten operators including QCD-penguin and
electroweak penguin
operators had to be  be considered which made the whole analysis rather
technical. The penguin operators have, however, no impact on the
coefficients $ C_1 $ and $ C_2 $ and also $ O(\alpha_{QED}) $
renormalization considered in \cite{BJL} can be neglected here. On the other
hand we are interested in the $ \mu$ dependence of $ a_1 $ and $ a_2 $
around $ \mu = O(m_b) $ and consequently we have to generalize the
numerical analysis of \cite{BJL}.

At this point it should be remarked that in the context of the
leading logarithmic approximation, the sensitivity of $a_2$ to
the precise values of $C_i$ has been emphasised in ref. \cite{K84}
long time ago. The expectation of K\"uhn and R\"uckl that higher
order QCD corrections should have an important impact on the
numerical values of $a_2$ turns out to be correct as we will
demonstrate explicitly below.

The main objectives of the present paper are:
\begin{itemize}
\item
The values of $ a_1(\mu) $ and $ a_2(\mu) $ beyond the leading
logarithmic approximation,
\item
The analysis of their $ \mu $  and $\Lms$ dependences,
\item
The analysis of their renormalization scheme dependence in
general terms, which we will illustrate here
by calculating $ a_i(\mu) $ in three renormalization schemes:
NDR, HV and DRED.
\end{itemize}

Since the $\mu$, $\Lms$ and the renormalization scheme dependences of
$a_i(\mu)$ are caused by the non-factorizable hard gluon contributions,
this analysis should give us some estimate of the expected departures
from factorization. It will also give us the answer whether, within the
theoretical uncertainties, the problem of the small value of $a_2$,
stressed by many authors in the past, can be avoided.

Our paper is organized as follows. In section 2  we give
a set of compact expressions
 for $ C_1(\mu) $
and $ C_2(\mu) $ which clearly exhibit the $ \mu $ and
renormalization scheme dependences. Subsequently in sections 3 and 4
we will critically analyse $a_i$ for B-decays and D-decays respectively.
Our main findings and conclusions are given in section 5.
\section{Master Formulae}
The coefficients $C_i(\mu)$ can be written as follows:
\begin{equation}\label{10}
C_1(\mu)=\frac{z_+(\mu)+z_-(\mu)}{2}
\qquad\qquad
C_2(\mu)=\frac{z_+(\mu)-z_-(\mu)}{2}
\end{equation}
where
\begin{equation}\label{11}
z_\pm(\mu)=\left[1+\frac{\alpha_s(\mu)}{4\pi}J_\pm\right]
      \left[\frac{\alpha_s(M_W)}{\alpha_s(\mu)}\right]^{d_\pm}
\left[1+\frac{\alpha_s(M_W)}{4\pi}(B_\pm-J_\pm)\right]
\end{equation}
with
\begin{equation}\label{12}
J_\pm=\frac{d_\pm}{\beta_0}\beta_1-\frac{\gamma^{(1)}_\pm}{2\beta_0}
\qquad\qquad
d_\pm=\frac{\gamma^{(0)}_\pm}{2\beta_0}
\end{equation}
\begin{equation}\label{13}
\gamma^{(0)}_\pm=\pm 2 (3\mp 1)
\qquad\quad
\beta_0=11-\frac{2}{3}f
\qquad\quad
\beta_1=102-\frac{38}{3}f
\end{equation}
\begin{equation}\label{14}
\gamma^{(1)}_{\pm}=\frac{3 \mp 1}{6}
\left[-21\pm\frac{4}{3}f-2\beta_0\kappa_\pm\right]
\end{equation}
\begin{equation}\label{15}
B_\pm=\frac{3 \mp 1}{6}\left[\pm 11+\kappa_\pm\right].
\end{equation}
Here we have introduced the parameter $\kappa_\pm$ which
distinguishes between various renormalization
schemes:
\begin{equation}\label{16}
\kappa_\pm = \left\{ \begin{array}{rc}
0 & (\rm{NDR})  \\
\mp 4 & (\rm{HV}) \\
\mp 6-3 & (\rm{DRED})
\end{array}\right.
\end{equation}

Thus $J_\pm$ in (\ref{12}) can also be written as
\begin{equation}\label{17}
J_\pm=(J_\pm)_{NDR}+\frac{3\mp 1}{6}\kappa_\pm
=(J_\pm)_{NDR}\pm\frac{\gamma^{(0)}_\pm}{12}\kappa_\pm
\end{equation}
Setting $\gamma_\pm^{(1)}$, $B_\pm$ and $\beta_1$ to zero
gives the leading logarithmic approximation \cite{LEE}. The
NLO corrections in the dimensional reduction scheme (DRED)
have been first considered in \cite{ALT}. The corresponding
calculations in the NDR scheme
 and in the HV scheme have been presented in \cite{WEISZ},
where the DRED-results of \cite{ALT} have been confirmed.
In writing (\ref{14}) we have incorporated
the $-2 \gamma^{(1)}_J$ correction
in the HV scheme resulting from the non-vanishing two--loop anomalous
dimension of the weak current. Similarly we have incorporated in
$\gamma^{(1)}_\pm$ a finite renormalization of $\alpha_s$ in the
case of the DRED scheme in order to work in all schemes with the usual
$\overline{MS}$ coupling \cite{BBDM}. For the latter we take
\begin{equation}\label{18}
\alpha_s(\mu)=\frac{4\pi}{\beta_0 \ln(\mu^2/\Lambda^2_{\overline{MS}})}
\left[1-\frac{\beta_1}{\beta^2_0}
\frac{\ln\ln(\mu^2/\Lambda^2_{\overline{MS}})}
{\ln(\mu^2/\Lambda^2_{\overline{MS}})}\right].
\end{equation}
The formulae given above
 depend on $f$, the number of active flavours. In the case of
B--decays $f=5$. According to the most recent world avarage
\cite{WEBER} we have:
\begin{equation}\label{19}
\alpha_s(M_Z)=0.117\pm0.007
\qquad\quad
\Lambda_{\overline{MS}}^{(5)}=(225\pm85)~MeV
\end{equation}
where the superscript stands for $f=5$.

In the case of D-decays the relevant scale is $\mu=O(m_c)$. In order
to calculate $C_i(\mu)$ for this case one has to evolve these
coefficients from $\mu=O(m_b)$ down to $\mu=O(m_c)$ in an effective
theory with $f=4$. Matching $\alpha_s^{(5)}(m_b)=\alpha_s^{(4)}(m_b)$
we find to a very good approximation $\Lms^{(4)}=(325\pm110)~MeV$.
Unfortunately the necessity to evolve $C_i(\mu)$ from $\mu=M_W$
down to $\mu=m_c$ in two different theories ($f=5$ and $f=4$) and
eventually with $f=3$ for $\mu< m_c$ makes the formulae
for $C_i(\mu)$ in D--decays rather complicated.
They can be found in \cite{BJL}.
Fortunately all these complications can be avoided by a simple trick,
which reproduces the results of \cite{BJL} to better than $0.5\%$.
In order to find $C_i(\mu)$ for $1~GeV\leq\mu\leq 2~GeV$ one can
simply use the master formulae given above with $\Lms^{(5)}$ replaced
by $\Lms^{(4)}$ and $f=4.15$. The latter "effective" value for $f$
allows to obtain a very good agreement with \cite{BJL}. The nice
feature of this method is that the $\mu$ and renormalization scheme
dependences of $C_i(\mu)$ can be studied in simple terms.

Returning to (\ref{11}) we note that $(B_\pm-J_\pm)$ is scheme independent.
The scheme dependence of $z_\pm(\mu)$ originates then entirely from
the scheme dependence of $J_\pm$ which has been explicitly shown
in (\ref{17}). We should stress that by the scheme dependence we always mean
the one related to the operator renormalization. The scheme for $\alpha_s$
is always $\overline{MS}$.
The scheme dependence present in the first factor in (\ref{11}) has
been removed in \cite{WEISZ} by multiplying $z_\pm(\mu)$ by
$(1-B_\pm \alpha_s(\mu)/4\pi)$ and the corresponding hadronic
matrix elements by $(1+B_\pm \alpha_s(\mu)/4\pi)$. Although this
procedure is valid in general, it is not useful in the case of
the factorization approach which precisely omitts the non-factorizable,
scheme dependent corrections such as $B_\pm$ or $J_\pm$ in the
hadronic matrix elements. Consequently in what follows we will work
with the true coefficients $C_i(\mu)$ of the operators $O_i$ as given
in (\ref{10}) and (\ref{11}).

In order to exhibit the $\mu$ dependence on the same footing as the
scheme dependence, it is useful to rewrite (\ref{11}) as follows:
\begin{equation}\label{20}
z_\pm(\mu)=\left[1+\frac{\alpha_s(m_b)}{4\pi} \tilde J_\pm(\mu)\right]
      \left[\frac{\alpha_s(M_W)}{\alpha_s(m_b)}\right]^{d_\pm}
\left[1+\frac{\alpha_s(M_W)}{4\pi}(B_\pm-J_\pm)\right]
\end{equation}
with
\begin{equation}\label{21}
\tilde J_\pm(\mu)=(J_\pm)_{NDR}\pm
\frac{\gamma^{(0)}_\pm}{12}\kappa_\pm
+\frac{\gamma^{(0)}_\pm}{2}\ln(\frac{\mu^2}{m^2_b})
\end{equation}
summarizing both the renormalization scheme dependence and the
$\mu$--dependence. Note that in the first parenthesis in (\ref{20})
we have
set $\alpha_s(\mu)=\alpha_s(m_b)$ as the difference in the
scales in this correction is still of a higher order.
We also note that the scheme and the $\mu$--dependent terms
are both proportional to $\gamma^{(0)}_\pm$. This implies that a
change of the renormalization scheme can be compensated by a change
in $\mu$. From (\ref{21}) we find generally
\begin{equation}\label{21a}
\mu_i^\pm=\mu_{NDR}\exp\left(\mp\frac{\kappa_\pm^{(i)}}{12}\right)
\end{equation}
where $i$ denotes a given scheme. From (\ref{16}) we have then
\begin{equation}\label{22}
\mu_{HV}=\mu_{NDR}\exp\left(\frac{1}{3}\right)
\qquad
\mu_{DRED}^{\pm}=
\mu_{NDR}\exp\left(\frac{2\pm 1}{4}\right)
\end{equation}
Evidently whereas the change in $\mu$ relating HV and NDR is the
same for $z_+$ and $z_-$ and consequently for $a_i(\mu)$ and
$C_i(\mu)$, the relation between NDR and DRED is more involved. In any
case $\mu_{HV}$ and $\mu_{DRED}^\pm$ are larger than $\mu_{NDR}$.
This discussion shows that a meaningful analysis of the $\mu$
dependence of $C_i(\mu)$ can only be made simultaneously with the
analysis of the scheme dependence.

Using (\ref{20}) and (\ref{21}) we can find the explicit dependence
of $a_i$ on $\mu$ and the renormalization scheme:
\begin{equation}\label{21c}
\Delta a_{1,2}(\mu)
=\frac{\alpha_s(m_b)}{3\pi}\left[F_+\mp F_-\right]
\ln(\frac{\mu^2}{m_b^2})+
\frac{\alpha_s(m_b)}{18\pi}\left[F_+\kappa_+\pm F_-\kappa_-\right]
\end{equation}
where $F_{\pm}$ denotes the product of the last two factors in
(\ref{20}) which are scheme independent.
For $m_b=4.8~GeV$, $\Lms^{(5)}=225\pm85~MeV$
we have $F_+=0.88\pm 0.01 $ and $F_-=1.28\pm 0.03$.
It is evident from (\ref{21c}) that
the $\mu$ and renormalization scheme dependences are much smaller
for $a_1$ than for $a_2$. We will verify this numerically below.

We have written all the formulae without invoking heavy quark
effective theory (HQET). It is sometimes stated in the literature
that for $\mu<m_b$ in the case of B-decays one {\it has to } switch
to HQET. In this case for $\mu<m_b$ the anomalous dimensions
 $\gamma_\pm$  differ from those given above \cite{GKMW}. We
should however stress that switching to HQET can be done at
any $\mu<m_b$ provided the logarithms $\ln(m_b/\mu)$ in
$\langle O_i \rangle$ do not become too large. Similar comments
apply to D-decays with respect to $\mu=m_c$. Of course the
coefficients $C_i$ calculated in HQET for $\mu<m_b$ are
different from the coefficients presented here. However the
corresponding matrix elements $\langle O_i \rangle$ in HQET are
also different so that the physical amplitudes remain unchanged.
Again, if factorization for $\langle O_i \rangle$ is used, it
matters to some extent at which $\mu$ the HQET is invoked.
For the range of $\mu$ considered here this turns out to be
inessential.

\begin{table}[thb]
\caption{The coefficient $C_1(\mu)$ for B-decays.}
\begin{center}
\begin{tabular}{|c|c|c|c||c|c|c||c|c|c|}
\hline
& \multicolumn{3}{c||}{$\Lms^{(5)}=140~MeV$} &
  \multicolumn{3}{c||}{$\Lms^{(5)}=225~MeV$} &
  \multicolumn{3}{c| }{$\Lms^{(5)}=310~MeV$} \\
\hline
$\mu [GeV]$ & NDR & HV & DRED & NDR &
HV & DRED & NDR & HV & DRED \\
\hline
\hline
4.0 & 1.074 & 1.092 & 1.073 & 1.086 & 1.107 & 1.086 &
1.096 & 1.120 & 1.097 \\
\hline
5.0 & 1.062 & 1.078 & 1.061 & 1.072 & 1.090 & 1.071 &
1.080 & 1.101 & 1.079 \\
\hline
6.0 & 1.054 & 1.069 & 1.052 & 1.062 & 1.079 & 1.060 &
1.068 & 1.087 & 1.067 \\
\hline
7.0 & 1.047 & 1.061 & 1.045 & 1.054 & 1.069 & 1.052 &
1.059 & 1.077 & 1.057 \\
\hline
8.0 & 1.042 & 1.055 & 1.039 & 1.047 & 1.062 & 1.045 &
1.052 & 1.069 & 1.050 \\
\hline
\end{tabular}
\end{center}
\end{table}

\begin{table}[thb]
\caption{The coefficient $C_2(\mu)$ for  B-decays.}
\begin{center}
\begin{tabular}{|c|c|c|c||c|c|c||c|c|c|}
\hline
& \multicolumn{3}{c||}{$\Lms^{(5)}=140~MeV$} &
  \multicolumn{3}{c||}{$\Lms^{(5)}=225~MeV$} &
  \multicolumn{3}{c| }{$\Lms^{(5)}=310~MeV$} \\
\hline
$\mu [GeV]$ & NDR & HV & DRED & NDR &
HV & DRED & NDR & HV & DRED \\
\hline
\hline
4.0 & --.175 & --.211 & --.216 & --.197 & --.239 & --.244 &
--.216 & --.264 & --.269 \\
\hline
5.0 & --.151 & --.184 & --.189 & --.169 & --.208 & --.213 &
--.185 & --.228 & --.233 \\
\hline
6.0 & --.133 & --.164 & --.169 & --.148 & --.184 & --.190 &
--.161 & --.201 & --.207 \\
\hline
7.0 & --.118 & --.148 & --.153 & --.132 & --.166 & --.171 &
--.143 & --.181 & --.186 \\
\hline
8.0 & --.106 & --.135 & --.140 & --.118 & --.151 & --.156 &
--.128 & --.164 & --.169 \\
\hline
\end{tabular}
\end{center}
\end{table}

\section{B-Decays}
The coefficients $C_i(\mu)$ are shown in tables
2 and 3 for different
$\mu$, $\Lms^{(5)}$ and the three renormalization schemes in question.
We include these results because they should be  useful independently
of the
factorization issue. The corresponding values for $a_i(\mu)$ are
given in tables 4 and 5. We observe:
\begin{itemize}
\item
the coefficient $a_1$ is very weakly dependent on $\mu$, $\Lms^{(5)}$
and the choice of the renormalization scheme. In the full range of
parameters considered we find:
\begin{equation}\label{35}
a_1=1.01\pm0.02
\end{equation}
in an excellent agreement with (\ref{8}). The weak dependence of
$a_1$ on the parameters considered can be
understood by inspecting (\ref{21c}).
\item
the coefficient $a_2$ depends much stronger on $\mu$, $\Lms^{(5)}$
and the choice of the renormalization scheme. Interestingly, for the
NDR scheme we find
\begin{equation}\label{36}
a_2^{NDR}=0.20\pm0.05
\end{equation}
which is in the ball park of the experimental findings in (\ref{8}).
Smaller
values are found for HV and DRED schemes:
\begin{equation}\label{37}
a_2^{HV}=0.16\pm0.05
\qquad
a_2^{DRED}=0.15\pm0.05
\end{equation}
\end{itemize}

\begin{table}[thb]
\caption{The coefficient $a_1(\mu)$ for B-decays.}
\begin{center}
\begin{tabular}{|c|c|c|c||c|c|c||c|c|c|}
\hline
& \multicolumn{3}{c||}{$\Lms^{(5)}=140~MeV$} &
  \multicolumn{3}{c||}{$\Lms^{(5)}=225~MeV$} &
  \multicolumn{3}{c| }{$\Lms^{(5)}=310~MeV$} \\
\hline
$\mu [GeV]$ & NDR & HV & DRED & NDR &
HV & DRED & NDR & HV & DRED \\
\hline
\hline
4.0 & 1.016 & 1.021 & 1.002 & 1.020 & 1.027 & 1.004 &
1.024 & 1.033 & 1.007 \\
\hline
5.0 & 1.012 & 1.017 & 0.998 & 1.015 & 1.021 & 1.000 &
1.018 & 1.025 & 1.002 \\
\hline
6.0 & 1.010 & 1.014 & 0.996 & 1.012 & 1.017 & 0.997 &
1.014 & 1.020 & 0.998 \\
\hline
7.0 & 1.008 & 1.011 & 0.994 & 1.010 & 1.014 & 0.995 &
1.012 & 1.017 & 0.995 \\
\hline
8.0 & 1.007 & 1.010 & 0.993 & 1.008 & 1.012 & 0.993 &
1.010 & 1.014 & 0.993 \\
\hline
\end{tabular}
\end{center}
\end{table}

\begin{table}[thb]
\caption{The coefficient $a_2(\mu)$ for B-decays.}
\begin{center}
\begin{tabular}{|c|c|c|c||c|c|c||c|c|c|}
\hline
& \multicolumn{3}{c||}{$\Lms^{(5)}=140~MeV$} &
  \multicolumn{3}{c||}{$\Lms^{(5)}=225~MeV$} &
  \multicolumn{3}{c| }{$\Lms^{(5)}=310~MeV$} \\
\hline
$\mu [GeV]$ & NDR & HV & DRED & NDR &
HV & DRED & NDR & HV & DRED \\
\hline
\hline
4.0 & 0.183 & 0.153 & 0.142 & 0.165 & 0.130 & 0.118 &
0.149 & 0.110 & 0.097 \\
\hline
5.0 & 0.203 & 0.175 & 0.164 & 0.188 & 0.156 & 0.144 &
0.175 & 0.139 & 0.127 \\
\hline
6.0 & 0.219 & 0.192 & 0.181 & 0.206 & 0.175 & 0.164 &
0.195 & 0.161 & 0.149 \\
\hline
7.0 & 0.231 & 0.205 & 0.195 & 0.220 & 0.191 & 0.179 &
0.210 & 0.178 & 0.166 \\
\hline
8.0 & 0.241 & 0.216 & 0.206 & 0.231 & 0.203 & 0.192 &
0.223 & 0.193 & 0.181 \\
\hline
\end{tabular}
\end{center}
\end{table}

This exercise shows that by including NLO QCD corrections and choosing
"appropriately" the renormalization scheme for the operators $O_i$,
one can achieve the agreement of the QCD factor $a_2$ in (\ref{6})
evaluated at $\mu=O(m_b)$ with the phenomenological findings. No high
scales
as found in the leading logarithmic approximation are necessary.
Moreover, as it is clear from (\ref{21a}), by choosing a scheme with
positve $\kappa_+$ and negative $\kappa_-$ even higher values for
$a_2$ at $\mu=m_b$ can be obtained.

In spite of the possibility of "fitting" the phenomenological
values for $a_2$ by choosing appropriately the renormalization scheme,
the sizable dependence of $a_2$ on $\mu$ and the renormalization
scheme is rather disturbing from the point of view of the factorization
approach. On the other hand it is interesting that within $2-3\%$
we find $a_1=1$
in the full range of the parameters considered.
We will return to these issues in
the final section.
\section{Charm Decays}
The phenomenological analyses of (\ref{1})-(\ref{3}) give in the case
of two-body D meson decays \cite{NEUBERT}:
\begin{equation}\label{38}
a_1\approx 1.2\pm0.10
\qquad
a_2\approx -0.5\pm 0.10
\end{equation}
The different sign of $a_2$ compared with the case of B-decays shows
that the structure of non-leptonic D decays differs considerably from
the one in B decays. Calculating $a_i$ according to our master formulae
for scales $1.0~GeV\leq \mu\leq 2.0~GeV$ we find that $a_1$
roughly agrees with
(\ref{38}). On the other hand as already found in the leading order
\cite{BAUER,RUCKL,NEUBERT}, the coefficient $a_2$ is generally
substantially smaller than its phenomenological value (\ref{38})
due to strong cancellation between $C_2$ and $C_1/3$. Only for
$\mu=1.0~GeV$, the largest $\Lms$ and HV and DRED schemes it is
possible to obtain $a_2$ within a factor of two from the value in
(\ref{38}). Otherwise one finds typically $a_2=O(0.1)$ and consequently
branching ratios for class II decays by an order of magnitude smaller
than the experimental branching ratios.

Because of these findings, a "new factorization" \cite{BAUER} approach
has been proposed in which the "1/N" terms in (\ref{6}) are discarded.
Some arguments for this modified approach can be given in the frameworks
of $1/N$ expansion \cite{RUCKL} and QCD sum rules \cite{BSQCD}.
 Yet "the rule of discarding $1/N$ terms" is certainly not established
both theoretically \cite{BLOK}
and phenomenologically \cite{STONE}.
Moreover as we already mentioned in the introduction,
it does not work for B decays giving wrong sign for $a_2$.
For completeness however we show in tables 6 and 7 the values of
$a_1=C_1$ and $a_2=C_2$ relevant for D-decays. We observe:
\begin{itemize}
\item
the coefficient $a_1$ is weakly dependent on the choice of
the renormalization scheme
for fixed $\mu$ and $\Lms^{(4)}$. The dependence on $\mu$ and $\Lms^{(4)}$
is sizable. In the full range of parameters we find
\begin{equation}\label{39}
a_1=1.31\pm 0.19
\end{equation}
which is compatible with phenomenology.
\item
the coefficient $a_2$ depends much stronger on the renormalization
scheme than $a_1$ and the dependence on $\mu$ and $\Lms^{(4)}$ is
really large.
Restricting the range of $\mu$ to $\mu=1.25\pm0.25~GeV$ we find
\begin{equation}\label{39a}
a_2^{NDR}=-0.47\pm 0.15
\qquad
a_2^{HV}\approx a_2^{DRED}\approx -0.60\pm0.22
\end{equation}
in the ball park of (\ref{38}).
\item
the dependences of $a_1$ and $a_2$ on the parameters considered are
stronger in the charm sector than in B decays because of the larger
QCD coupling involved.
\end{itemize}

\begin{table}[thb]
\caption{The coefficient $C_1(\mu)$ for D-decays.}
\begin{center}
\begin{tabular}{|c|c|c|c||c|c|c||c|c|c|}
\hline
& \multicolumn{3}{c||}{$\Lms^{(4)}=215~MeV$} &
  \multicolumn{3}{c||}{$\Lms^{(4)}=325~MeV$} &
  \multicolumn{3}{c| }{$\Lms^{(4)}=435~MeV$} \\
\hline
$\mu [GeV]$ & NDR & HV & DRED & NDR &
HV & DRED & NDR & HV & DRED \\
\hline
\hline
1.00 & 1.208 & 1.259 & 1.224 & 1.275 & 1.358 & 1.309 &
1.363 & 1.506 & 1.432 \\
\hline
1.25 & 1.174 & 1.216 & 1.185 & 1.221 & 1.282 & 1.242 &
1.277 & 1.367 & 1.314 \\
\hline
1.50 & 1.152 & 1.187 & 1.160 & 1.188 & 1.237 & 1.203 &
1.228 & 1.296 & 1.252 \\
\hline
1.75 & 1.136 & 1.167 & 1.142 & 1.165 & 1.207 & 1.176 &
1.196 & 1.252 & 1.214 \\
\hline
2.00 & 1.123 & 1.152 & 1.128 & 1.148 & 1.185 & 1.156 &
1.174 & 1.221 & 1.187 \\
\hline
\end{tabular}
\end{center}
\end{table}

\begin{table}[thb]
\caption{The coefficient $C_2(\mu)$ for D-decays.}
\begin{center}
\begin{tabular}{|c|c|c|c||c|c|c||c|c|c|}
\hline
& \multicolumn{3}{c||}{$\Lms^{(4)}=215~MeV$} &
  \multicolumn{3}{c||}{$\Lms^{(4)}=325~MeV$} &
  \multicolumn{3}{c| }{$\Lms^{(4)}=435~MeV$} \\
\hline
$\mu [GeV]$ & NDR & HV & DRED & NDR &
HV & DRED & NDR & HV & DRED \\
\hline
\hline
1.00 & --.410 & --.491 & --.492 & --.510 & --.631 & --.630 &
--.632 & --.825 & --.815 \\
\hline
1.25 & --.356 & --.424 & --.427 & --.430 & --.523 & --.525 &
--.512 & --.642 & --.640 \\
\hline
1.50 & --.319 & --.379 & --.383 & --.378 & --.457 & --.459 &
--.439 & --.543 & --.543 \\
\hline
1.75 & --.291 & --.346 & --.350 & --.340 & --.410 & --.414 &
--.390 & --.478 & --.480 \\
\hline
2.00 & --.269 & --.320 & --.324 & --.311 & --.375 & --.379 &
--.353 & --.431 & --.435 \\
\hline
\end{tabular}
\end{center}
\end{table}
\section{Final Remarks}
We have calculated the QCD factors $a_1$ and $a_2$,
entering the tests of factorization in non-leptonic heavy meson decays,
beyond the leading logarithmic approximation. In particular we have
pointed out that $a_i$ in QCD depend not only on $\mu$ and $\Lms$,
but also on the renormalization scheme for the operators. The latter
dependence precludes a unique determination of the factorization
scale $\mu_F$, if such a scale exists at all, at which the factorization
approach would give results identical to QCD. For instance going from
the NDR scheme to the HV scheme is equivalent, in the case of
current-current
operators $O_i$, to a change of $\mu_F$ by $40\%$. Simultaneously
we would like to emphasize
the strong dependence of a possible $\mu_F$ on
$\Lms$. The latter uncertainty can however be considerably reduced in
the future by reducing the uncertainty in $\Lms$ extracted from high
energy processes. The NLO calculations of $a_i$ and $C_i$ presented
here, allow a meaningful use of $\Lms$, extracted from high energy
processes, in the non-leptonic decays in question.

On the phenomenological side the following results are in our opinion
interesting. In the simplest renormalization scheme with anti-commuting
$\gamma_5$ (NDR), $\Lms^{(5)}=(225\pm85)~MeV$ and $\mu=6\pm2~GeV$,
we find in the case of B-decays
\begin{equation}\label{40}
a_1^{NDR}=1.02\pm0.01
\qquad
a_2^{NDR}=0.20\pm0.05
\end{equation}
which are in the ball park of the results of phenomenological analyses.
In particular, the inclusion of NLO corrections in the NDR scheme
appears to "solve" the problem of the small value of $a_2$ obtained
in the leading order.

In the case of D-decays, $\Lms^{(4)}=(325\pm110)~MeV$,
 $\mu=1.25\pm0.25~GeV$ and using the "new factorization" approach we
find
\begin{equation}\label{41}
a_1^{NDR}=1.26\pm0.10
\qquad
a_2^{NDR}=-0.47\pm0.15
\end{equation}
again in the ball park of phenomenological analyses. The standard
factorization gives for D-decays $a_1^{NDR}\approx 1.10\pm0.05$
and $a_2^{NDR}\approx-0.06\pm0.12$ for the same range of parameters.
The result for $a_2$ is phenomenologically inacceptable.

We have also stressed that similar results for $a_1$ in B-decays
are obtained in HV and DRED schemes. Moreover the very weak dependence
of $a_1$ on $\mu$ and $\Lms$ indicates that $a_1$ is predicted to be
close to unity in agreement with phenomenology of factorization.
However the $\mu$, $\Lms$ and scheme dependences of $a_2$
for  B decays and in particular for D decays
are rather sizable.

In our opinion the failure of the usual factorization approach in
D decays and the strong dependence of $a_2$ on $\mu$, $\Lms$ and
the choice of the renormalization scheme indicate
 that non-factorizable contributions must
play generally an important role in heavy meson non-leptonic decays
if QCD is
the correct description of these decays.
In K meson decays the non-factorizable contributions are known to
be very important anyway \cite{BLOK,NONFACT}.
Consequently we expect that,
when the experimental data improves, sizable departures from factorization
should become visible in particular in decays belonging to class II.
An exception could be the class I in B decays where an accidental
approximate cancellations of $\mu$ and renormalization scheme dependences
takes place in $a_1$. It should however be stressed that the stability
of $a_1$ with respect to changes of $\mu$ and the renormalization
scheme is only a necessary condition for an "effective" validity of
factorization in class I decays. It certainly does not imply that
factorization of matrix elements indeed takes place.

In spite of these critical remarks the tests of factorization in
non-leptonic decays are important because the patterns of the expected
departures from factorization will teach us about the non-factorizable
contributions. Recent discussion of such contributions
can be found in \cite{RUCKL94}.
In this connection, once the data and the models for
formfactors improve, it would be useful to investigate in detail how the
phenomenologically extracted parameters $a_1$ and $a_2$ depend on the
decay channel considered.

I would like to thank Gerhard Buchalla and Robert Fleischer for
critical reading of the manuscript. I also thank Reinhold R\"uckl
for a discussion related to his work.

\vfill\eject


\begin{thebibliography}{99}
\bibitem{FEYNMAN}
J. Schwinger, {\sl Phys. Rev. Lett.} {\bf 12} (1964) 630;
R.P. Feynman, in {\it Symmetries in Particle Physics}, ed. A. Zichichi,
Acad. Press 1965, p.167; O. Haan and B. Stech,
{\sl Nucl. Phys.} {\bf B 22}  (1970) 448.
\bibitem{STECHF}
D. Fakirov and B. Stech, {\sl Nucl. Phys.} {\bf B 133}  (1978) 315;
L.L. Chau, Phys. Rep. {\bf B 95} (1983) 1.
\bibitem{BAUER}
M. Wirbel, B. Stech and M. Bauer, {\sl Z. Phys.} {\bf C 29} (1985) 637.
M. Bauer, B. Stech and M. Wirbel, {\sl Z. Phys.} {\bf C 34} (1987) 103.
\bibitem{NEUBERT}
M. Neubert, V. Rieckert, B. Stech and Q.P. Xu, in "Heavy Flavours",
 eds. A.J. Buras and M. Lindner (World Scientific, Singapore, 1992),
p. 286;
\bibitem{BJL}
A.J. Buras, M. Jamin, and M.E. Lautenbacher,
{\sl Nucl. Phys.} {\bf B 408} (1993) 209.
\bibitem{ROME}
M. Ciuchini, E. Franco, G. Martinelli and L. Reina,
 {\sl Phys. Lett.} {\bf B 301} (1993) 263;
{\sl Nucl. Phys.} {\bf B 415} (1994) 403.
\bibitem{BJLW}
A.J. Buras, M. Jamin, M.E. Lautenbacher, and P.H. Weisz,
{\sl Nucl. Phys.} {\bf B 370} (1992) 69.
\bibitem{BJORKEN}
J.D. Bjorken,
{\sl Nucl. Phys.} {\bf B } (Proc. Suppl.) 11 (1989) 325;
SLAC-PUB-5389.
\bibitem{DUGAN}
M.J. Dugan and B. Grinstein,
{\sl Phys. Lett.} {\bf B 255} (1991) 583.
\bibitem{ISGUR}
C. Reader and N. Isgur,
{\sl Phys. Rev.} {\bf D 47} (1993) 1007.
\bibitem{DEANDREA}
A. Deandrea, N. Di Bartolomeo, R. Gatto and G. Nardulli
 {\sl Phys. Lett.} {\bf B 318} (1993) 549.
 \bibitem{GOURDIN}
M. Gourdin, A.N. Kamal, Y.Y. Keum and X.Y. Pham,
 {\sl Phys. Lett.} {\bf B 333} (1994) 507.
\bibitem{BROWDER}
T.E. Browder, K. Honscheid and S. Playfer,
in "B Decays II", ed. S. Stone, World Scientific (1994).
\bibitem{SCHUBERT}
K. Honscheid, K.R. Schubert and  R. Waldi, OHSTPY-HEP-E-93-017.
\bibitem{CLEO}
M.S. Alam et al. (CLEO Collaboration),
{\sl Phys. Rev.} {\bf D 50} (1994) 43.
\bibitem{KAMAL}
A.N. Kamal and T.N. Pham,
{\sl Phys. Rev.} {\bf D 50} (1994) 395.
\bibitem{STONE90}
D. Bortoletto and S. Stone,  {\sl Phys. Rev. Lett.} {\bf 65} (1990) 2951.
\bibitem{RUCKL}
A.J. Buras, J.M. G\'erard and R. R\"uckl,
{\sl Nucl. Phys.} {\bf B 268} (1986) 16.
\bibitem{BIGI}
I. Bigi, B. Blok, M. Shifman, N. Uraltsev and A. Vainshtein,
in "B Decays II", ed. S. Stone, World Scientific (1994).
\bibitem{RUCKL94}
A. Khodjamirian and R. R\"uckl, MPI-PhT/94-26, LMU 05/94.
\bibitem{LEE}
G. Altarelli and L. Maiani, {\sl Phys. Lett.} {\bf B 52} (1974) 351;
M.K. Gaillard and B.W. Lee, {\sl Phys. Rev. Lett.} {\bf 33} (1974) 108.
\bibitem{BSQCD}
B. Blok and M.A. Shifman,
{\sl Sov. Journ. Nucl. Phys.} {\bf 45} (1987) 135; 301; 522.
\bibitem{BLOK}
M.A. Shifman, {\sl Nucl. Phys.} {\bf B 388} (1992) 346;
 B. Blok and M.A. Shifman,
{\sl Nucl. Phys.} {\bf B 389} (1993) 534;
{\sl Nucl. Phys.} {\bf B 399} (1993) 441, 459.
\bibitem{ALT}
G. Altarelli, G. Curci, G. Martinelli and S. Petrarca,
{\sl Nucl. Phys.} {\bf B 187} (1981) 461.
\bibitem{WEISZ}
A.J. Buras and P.H. Weisz,
{\sl Nucl. Phys.} {\bf B 333} (1990) 66.
\bibitem{K84}
J.H. K\"uhn and R. R\"uckl,
{\sl Phys. Lett.} {\bf B 135} (1984) 477;
\bibitem{BBDM}
W.A. Bardeen, A.J. Buras, D.W. Duke and T. Muta,
{\sl Phys. Rev.} {\bf D 18} (1978) 3998.
\bibitem{WEBER}
S. Bethke, QCD 94 conference, Montpellier, 7-13 July 1994;
B.R. Webber, International Conference on High Energy Physics,
Glasgow, 20-27 July 1994.
\bibitem{GKMW}
B. Grinstein, W.Kilian, T. Mannel and M. Wise,
{\sl Nucl. Phys.} {\bf B 363} (1991) 19.
\bibitem{STONE}
S. Stone, in Heavy Flavours, eds. A.J Buras and M. Lindner,
World Scientific (1992), p.334.
\bibitem{NONFACT}
W.A. Bardeen, A.J. Buras and J.-M. G\'erard,
{\sl Phys. Lett.} {\bf B 192} (1987) 138;
A. Pich and E. de Rafael,
{\sl Nucl. Phys.} {\bf B 358} (1991) 311.
M. Neubert and B. Stech,
{\sl Phys. Rev.} {\bf D 44} (1991) 775.
\end{thebibliography}
\end{document}